\documentclass[superscriptaddress,secnumarabic,nobibnotes,aps,prd,showkeys,showpacs,nofootinbib,preprint]{revtex4}

\usepackage{graphicx}
\usepackage{float}
\usepackage{bm}
\usepackage{amsmath}
\usepackage{amsfonts}
\usepackage{amssymb}
\usepackage{epstopdf}
\usepackage{natbib}%
\newcommand{\bee}{\begin{equation}}
\newcommand{\eee}{\end{equation}}
\newcommand{\eaa}{\end{eqnarray}}
\newcommand{\baa}{\begin{eqnarray}}
\def\ni{\noindent}

\usepackage{color}

\begin{document}

\title{Loop Quantum Gravity Immirzi parameter \\ and the Kaniadakis statistics}

\author{Everton M. C. Abreu}\email{evertonabreu@ufrrj.br}
\affiliation{Grupo de F\' isica Te\'orica e F\' isica Matem\'atica, Departamento de F\'{i}sica, Universidade Federal Rural do Rio de Janeiro, 23890-971, Serop\'edica, RJ, Brazil}
\affiliation{Departamento de F\'{i}sica, Universidade Federal de Juiz de Fora, 36036-330, Juiz de Fora, MG, Brazil}
\author{Jorge Ananias Neto}\email{jorge@fisica.ufjf.br}
\affiliation{Departamento de F\'{i}sica, Universidade Federal de Juiz de Fora, 36036-330, Juiz de Fora, MG, Brazil}
\author{Albert C. R. Mendes}\email{albert@fisica.ufjf.br}
\affiliation{Departamento de F\'{i}sica, Universidade Federal de Juiz de Fora, 36036-330, Juiz de Fora, MG, Brazil}
\author{Rodrigo M. de Paula}\email{rmpaula@fisica.ufjf.br}
\affiliation{Departamento de F\'{i}sica, Universidade Federal de Juiz de Fora, 36036-330, Juiz de Fora, MG, Brazil}


\begin{abstract}
\noindent In this letter we have shown that a possible connection between the LQG Immirzi parameter and the area of a punctured surface can emerge depending on the thermostatistics theory previously chosen. Starting from the Boltzmann-Gibbs entropy, the Immirzi parameter can be reobtained. Using the Kaniadakis statistics, which is an important non-Gaussian statistics, we have derived a new relation between the Immirzi parameter, the kappa parameter and the area of a punctured surface. After that, we have compared our result with the Immirzi parameter previously obtained in the literature within the context of Tsallis' statistics.   We have demonstrated in an exact way that the LQG Immirzi parameter can also be used to compare both Kaniadakis and Tsallis statics. 

\end{abstract}
\date{\today}

\maketitle

 Loop quantum gravity (LQG)\cite{lqgr} exhibits quantum operators for area and volume that have discrete spectra.   It is a canonical quantization of the classical field.   Its formalism uses spin networks as being its Hilbert space.
Although these results are very important, both of them present difficulties which are the existence, in principle, of a free dimensionless parameter which it is called Immirzi parameter \cite{immi}. In a few words we can say that the Immirzi parameter provides the size of a quantum of area in Planck units. The Immirzi parameter correspond to a one-parameter set of canonical transformation relative to the canonical variables of the classical theory.   Namely, for every $\gamma$, the classical equations of motion  of general relativity are irrefutable.   At the quantum level, we can say that $\gamma$ corresponds to a quantization ambiguity of the theory.   There is a real quantum theory for each positive and real $\gamma$ \cite{maj}.
It can be computed by counting the number of microstates in LQG.   The quantum geometry of a cross-section of a black hole horizon in LQG is described by a topological two-sphere with defects, usually called punctures, carrying ``spin" quantum numbers endowed by the edges of the spin network that represent the bulk quantum geometry.
One way to compute the Immirzi parameter, solving the problem mentioned above, can be with the help of the Bekenstein-Hawking entropy area law (BHEAL) \cite{od}. The procedure usually adopted is that BHEAL has its origin in the Boltzmann-Gibbs (BG) entropy. In this letter we will perform a modification of the method mentioned above in such a way that the BHEAL is now a consequence of an important non-Gaussian statistics, namely, the Kaniadakis statistics \cite{kani1}.
 
The Hilbert space of LQG is formed by the spin networks. In the case of gravity, spin networks are graphs whose edges are labeled by representations of the $SU(2)$ gauge group. If a surface is intersected by an edge of a spin network carrying the label $j$, then the quantum area of the black hole is
\begin{eqnarray}
\label{a1}
a(j)=8 \pi l_p^2 \gamma \sqrt{j(j+1)}\,,
\end{eqnarray}

\ni where $l_p$ is the Planck length, $\gamma$ is the  Immirzi parameter and $j$, the spin quantum numbers $j_1,\ldots,j_N$ on $N$ punctures, can take values like $0, \frac{1}{2}, 1, \frac{3}{2}, ...$ i.e., the representation of $SU(2)$ group.   These labels represent the edges of the spin networks. From Eq. (\ref{a1}) we can conclude that the area of a black-hole horizon
can be formed by a large number of spin network edges puncturing the surface.

The well known Kaniadakis statistics, also refereed as $\kappa$-statistics, analogously to Tsallis thermostatistics model \cite{tsallis}, generalizes the usual BG statistics initially by introducing both the $\kappa$-exponential and $\kappa$-logarithm defined respectively by
\begin{eqnarray}
\label{expk}
exp_\kappa(f)=\Big( \sqrt{1+\kappa^2 f^2}+\kappa f \Big)^\frac{1}{\kappa}\,,
\end{eqnarray}
\begin{eqnarray}
\label{logk}
\ln_\kappa(f)=\frac{f^\kappa-f^{-\kappa}}{2\kappa}\,,
\end{eqnarray}

\ni and the following property can be satisfied, namely
\begin{eqnarray}
\ln_\kappa\Big(exp_\kappa(f)\Big)=exp_\kappa\Big(\ln_\kappa(f)\Big)\equiv f\,.
\end{eqnarray}

\ni From Eqs. (\ref{expk}) and (\ref{logk}) we can notice that the $\kappa$-parameter twists the standard definitions of the exponential and logarithm functions.

The $\kappa$-entropy, connected to this $\kappa$-framework, can be written as
\begin{eqnarray}
S_\kappa=- k_B \sum_i^W  \,\frac{p_i^{1+\kappa}-p_i^{1-\kappa}}{2\kappa}\,,
\end{eqnarray}

\ni where $p_i$ is the probability of the system to be in a microstate, $W$ is the total number of configurations and $\kappa$ is known as the kappa parameter.   At the limit $\kappa \rightarrow 0$ we must  recover the BG entropy. It is relevant to comment here that the $\kappa$-entropy satisfies the properties concerning concavity, additivity and extensivity. The $\kappa$-statistics has thrived when applied in many experimental scenarios. As an example we can cite cosmic rays  \cite{Kanisca1,Kanisca2}, cosmic effects  \cite{aabn-1}, astrophysical models \cite{aabn-2}, quark-gluon plasma  \cite{Tewe}, kinetic models describing a gas of interacting atoms and photons  \cite{Ross} and financial models  \cite{RBJ}.

Using the microcanonical ensemble definition, where all the states have the same probability, Kaniadakis' entropy reduces to \cite{kani1,Kanisca1}
\begin{eqnarray}
\label{microk}
S_\kappa=k_B\, \frac{W^\kappa-W^{-\kappa}}{2\kappa}\;,
\end{eqnarray}

\ni where at the limit $\kappa \rightarrow 0$, as we have already mentioned,  we must recover the usual BG entropy formula, i.e., $S=k_B\, \ln {W}$.

In order to derive the dependence of the Immirzi parameter \cite{od}, we will consider at first the BG statistics. The number of configurations (microstates) in a punctured surface is given by

\begin{eqnarray}
\label{nconf}
W=\prod_{n=1}^{N} (2 j_n+1)\,,
\end{eqnarray}

\ni where the term $(2 j_n+1)$ in (\ref{nconf}) is the multiplicity of the state $j$. It can be shown that the most important configurations that contribute in Eq.(\ref{nconf}) are those in which the lowest possible value for the spin dominates. Denoting the lowest value of spin as $j_{min}$,  then from Eq. (\ref{nconf}) we have 

\begin{eqnarray}
\label{micron}
W=(2 j_{min}+1)^N\,.
\end{eqnarray}

\ni From Eq. (\ref{a1}) we can obtain the number of punctures in a surface of area $A$ which is given by
\begin{eqnarray}
\label{np}
N=\frac{A}{a(j_{min})}=\frac{A}{8 \pi \, l_p^2 \gamma \sqrt{j_{min} ( j_{min}+1)}}\,.
\end{eqnarray}

\ni In BG statistics the entropy is given by $S=k_B\, \ln W$. Using Eq. (\ref{micron}) we have
\begin{eqnarray}
\label{bgl}
S=k_B N \ln (2 j_{min}+1)\,.
\end{eqnarray}

\ni Equating  the BHEAL that says that $S=k_B\,A/4l_p^2$  with Eq.(\ref{bgl}), using Eq. \eqref{np} and that 
$\,j_{min}=1/2$ (the most important contribution, statistically speaking, since it is the lowest nontrivial representation of $SU(2)$) we obtain the Immirzi parameter  
\begin{eqnarray}
\label{nbg}
\gamma=\frac{\ln 2}{\pi \sqrt{3}}\,.
\end{eqnarray}

We can generalize the procedure adopted above and, consequently, to obtain a more general relation for the Immirzi parameter. To complete the task (and we can see a similar way described in references \cite{maj} and \cite{sag}), we will use Kaniadakis entropy in the microcanonical ensemble. Using Eq. (\ref{micron}) and Kaniadakis entropy, Eq. (\ref{microk}), we have

\begin{eqnarray}
\label{eqk}
k_B \,\frac{(2 j_{min}+1)^{N\kappa}-(2 j_{min}+1)^{-N\kappa}}{2\kappa}=k_B \frac{A}{4l_p^2}
\end{eqnarray}

\begin{eqnarray}
\label{npk}
\Longrightarrow N=\frac{\ln\left[\frac{\kappa A}{4 l_p^2} + \sqrt{\frac{\kappa^2 A^2}{16 l_p^4} +1 } \right]}{\kappa \ln (2 j_{min}+1)}\,\,.
\end{eqnarray}

\ni Substituting $j_{min}=1/2$ in Eq.(\ref{npk}) then the number of punctures is given by

\begin{eqnarray}
\label{npk2}
N=\frac{\ln\left[\frac{\kappa A}{4 l_p^2} + \sqrt{\frac{\kappa^2 A^2}{16 l_p^4} +1 } \right]}{\kappa \ln 2 }\,\,.
\end{eqnarray}

\ni It is important to mention that, for a given value of $A$, the number of punctures $N$ in Eq. (\ref{npk2}) is now a function of the kappa parameter $\kappa$. From Eqs. (\ref{np}) and (\ref{npk2}) we can derive the kappa version of the Immirzi parameter which is given by

\begin{eqnarray}
\label{ki}
\gamma_\kappa= \frac{\ln 2}{\pi \sqrt{3}} \,\, \frac{ \frac{\kappa\, A}{4 l_p^2}}{\ln\left[\frac{\kappa A}{4 l_p^2} + \sqrt{1+\frac{\kappa^2  A^2 }{16 l_p^4}}\right]\;}\,.
\end{eqnarray}

\ni From Eq. (\ref{ki}) we can observe that in the limit $\kappa \rightarrow 0$ we recover the Immirzi parameter $\gamma= \ln 2/(\pi \sqrt{3})$. In Fig. 1 we have plotted Eq. (\ref{ki}) as a function of $\kappa$. In the same figure we have also plotted the Immirzi parameter derived in the context of Tsallis statistics which was obtained by Majhi \cite{maj} and it is written as 

\begin{eqnarray}
\label{ti}
\gamma_q= \frac{\ln 2}{\pi \sqrt{3}} \,\, \frac{\frac{(1-q)A}{4l_p^2}}{\ln[1+ (1-q)\frac{A}{4l_p^2}]}\nonumber\\
=-\frac{\ln 2}{\pi \sqrt{3}} \,\, \frac{\frac{\kappa A}{4l_p^2}}
{\ln[1-\frac{\kappa A}{4l_p^2}]}\,.
\end{eqnarray}

\ni In order to compare Eqs. (\ref{ki}) and (\ref{ti}) here we have used in Eq. (\ref{ti}) the relation \cite{bss}
\begin{eqnarray}
\label{rkq}
\kappa=q-1 \,,
\end{eqnarray}

\ni where $q$ is the Tsallis nonextensive parameter. We have also considered that $\,A=10^6 \,l_p^2 \,$ in Eqs. (\ref{ki}) and (\ref{ti}), in Fig. (1). 

From Fig. 1 we can observe that the difference between both the Tsallis and Kaniadakis computations of the Immirzi parameter are more pronounced for large kappa module parameter (for $A=10^6$).  In Fig 2, we can see the $\gamma$ behavior concerning a variable $A$.  In Fig. 3, we can see a 3D version of Fig. 1.  From these results we can see that using Kaniadakis statistics the values of $\gamma$ are real and positive, confirming the results of \cite{maj} for Tsallis statistics in LQG.  From Fig. 1 we can realize also that, in the case of a black hole described in LQG, since $\gamma$ increases in magnitude as the coupling between the horizon and the bulk becomes stronger \cite{maj}, analyzed from Kaniadakis formulation, the growth is slower than in Tsallis point of view.

In \cite{maj}, the author has demonstrated  that the parameter $q$ is a function of $A$.   Since the $q$-parameter characterizes the coupling effect between the horizon and the bulk, this coupling strength is determined by $A$.  From Eq. \eqref{ki}, we have shown that the $\kappa$-parameter is also a function of $A$.   In \cite{bss}, the authors have shown, using the polytropic index, that, although Tsallis and Kaniadakis show analogous behaviors, the astrophysical limit provides different limits for both.  In Fig. 1 and 3, we can see exactly this difference since we have shown precisely both behaviors concerning LQG through the use of Immirzi parameter.

\begin{figure}[H]
\begin{center}
\label{immik}
\includegraphics[width=3in, height=3in]{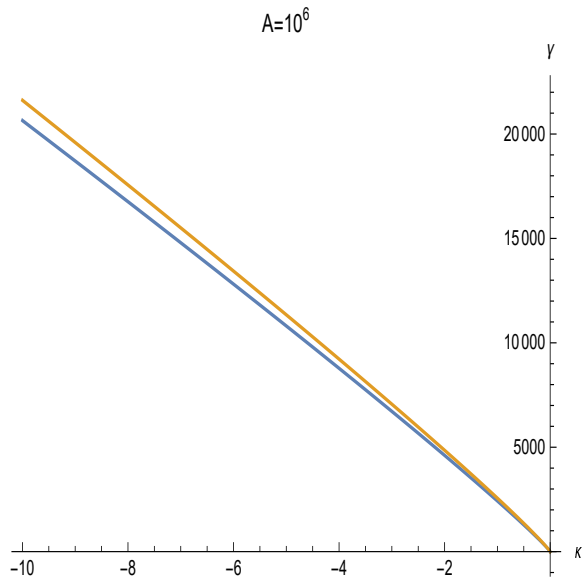}
\caption{The Immirzi parameter as a function of $\kappa$'s. The blue curve (below) is relative to Kaniadakis statistics and the other one, Tsallis statistics. }
\end{center}
\end{figure}

\begin{figure}[H]
\begin{center}
\label{immik}
\includegraphics[width=3in, height=3in]{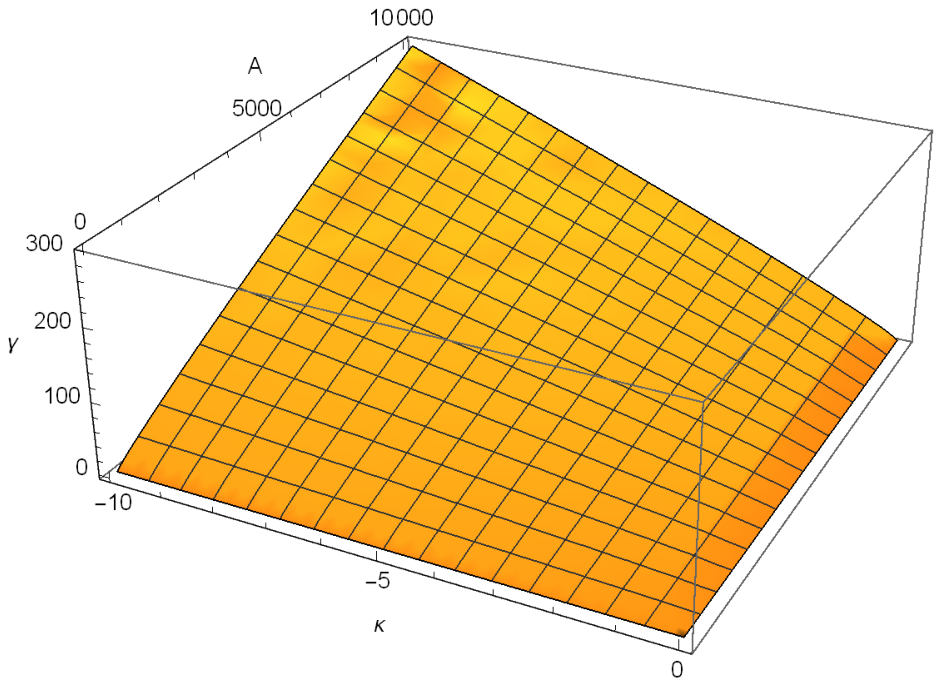}
\caption{Tri-dimensional Immirzi parameter ($\gamma$) curve relative to Kaniadakis statistics as a function of $\kappa$ and the normalized area, namely, $A \rightarrow A/l^2_P$.}
\end{center}
\end{figure}

\begin{figure}[H]
\begin{center}
\label{immik}
\includegraphics[width=3in, height=3in]{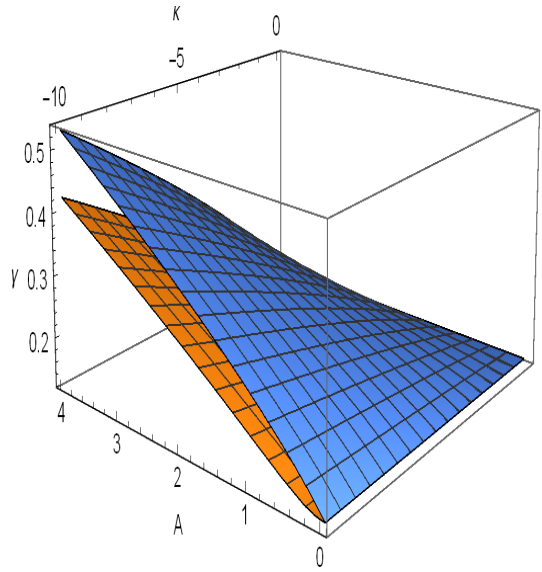}
\caption{Tri-dimensional Immirzi parameter ($\gamma$) curve relative to both Kaniadakis (yellow) and Tsallis statistics as a function of $\kappa$ and the normalized area, namely, $A \rightarrow A/l^2_P$.}
\end{center}
\end{figure}

To sum up, in this work we have investigated the effect of the Kaniadakis statistics in the framework of LQG, more specifically in the context of Immirzi parameter. In principle, the Immirzi parameter is arbitrary and one way of determining it is to use BHEAL. From the analytical viewpoint, we have derived in the framework of Kaniadakis' statistics the Immirzi parameter, Eq. (\ref{ki}), as a function of the kappa parameter and the area of the punctured surface. In the limit $\kappa \rightarrow 0$, where BG must be recovered, we have reobtained the Immirzi parameter 
$\gamma=\ln 2/(\pi \sqrt{3})$. After that, we have compared our result with that obtained in the Tsallis statistic, Eq. (\ref{ti}). We can observe in Fig. 1 that the Immirzi parameters derived from both Kaniadakis and Tsallis statistics present similar behavior. Therefore our result, together with others in the literature \cite{nossos-1,nossos-2}, can possibly indicates that if we consider Tsallis' statistics as an important possible extension of BG entropy thus Kaniadakis' statistics should also be considered as a viable theory in order to extend the BG theory.

\section*{Acknowledgments}

\ni The authors thank CNPq (Conselho Nacional de Desenvolvimento Cient\' ifico e Tecnol\'ogico), Brazilian scientific support federal agency, for partial financial support, Grants numbers 302155/2015-5 (E.M.C.A.) and 303140/2017-8 (J.A.N.). E.M.C.A. thanks the hospitality of Theoretical Physics Department at Federal University of Rio de Janeiro (UFRJ), where part of this work was carried out.

\end{document}